\documentclass[final,3p,times]{elsarticle}

\usepackage{graphicx}% Include figure files
\usepackage{lineno}
\usepackage{amsmath}
\usepackage{ulem}
\usepackage{color}

\usepackage{bm}% bold math

\usepackage{soul}

\makeatletter
\def\ps@pprintTitle{%
  \let\@oddhead\@empty
  \let\@evenhead\@empty
  \let\@oddfoot\@empty
  \let\@evenfoot\@oddfoot
}
\makeatother
\begin{document}

\title{Homophilic Effects on Economic Inequality: A Dynamic Network Agent-Based Model}

\author[1]{Gustavo L. Kohlrausch}
\ead{gustavo.luis.kohlrausch@gmail.com}
\author[2]{Thiago Dias}
\author[1]{Sebastian Gonçalves}
\affiliation[1]{Instituto de Física, Universidade Federal do Rio Grande do Sul,  91501-970 Porto Alegre, RS, Brazil}

\affiliation[2]{Universidade Tecnológica Federal do Paraná, 85660-000 Dois Vizinhos, PR, Brazil }

\begin{abstract}
Wealth transactions are central to economic activity, and their particularities shape macroeconomic outcomes. We propose an agent-based model to investigate how homophily influences economic inequality. The model simulates wealth exchanges in a dynamic network composed of two groups, $A$ and $B$, 
differentiated by a homophily parameter $\delta$, which increases intragroup connections within $A$. 
Economic interactions alternate between conservative wealth exchanges and connection rewiring, 
both influenced by agents' wealth and $\delta$. We examine economic and network dynamics under 
varying levels of social protection $f$, which favor poorer agents in transactions. At low $f$, results reveal high inequality and link concentration, with $\delta$ impacting only transient dynamics. At high $f$, homophily becomes an economic advantage, as increasing $\delta$ directs wealth flow to group $A$. 
However, since this flow benefits the wealthiest agents, it simultaneously exacerbates internal inequality within the group. These findings show that homophily is a significant driver of inequality, directing wealth towards the homophilous group and worsening internal disparities. 
\end{abstract}

\maketitle

\section{Introduction}

Economic inequality represents an deep-rooted problem in an increasing globalized world, which presents a rising trend over the years \cite{Alvaredo2018, capitalXXI, saez2016wealth,WorldBank,IncomeChancel}. 
Despite geographical, cultural, and historical differences,  wealth and income distributions among individuals in various countries demonstrates a universal pattern, which is well-fitted by a two pattern division \cite{Chakrabarti2013,Yakovenko2001, Ludwig2022}, where the upper tail, typically accounting less than $5\%$, displays a power law while the majority of income follows a Gamma distribution  \cite{Clementi2005}. 
This universality indicates a set of fundamental mechanisms that lead to economic inequalities regardless of the countries. Understanding the basic mechanisms that creates such stratification is essential to devise efficient ways to avoid or reduce it. The pursuit of understanding such mechanisms garnered the attention of econophysicists, who merging economic theory and statistical pysics tools \cite{slanina2013,jusup2022social,kutner2019}, utilize first-principles calculations and agent-based simulations to reproduce  macroeconomic quantities \cite{Chakrabarti2013, Benhur, Iglesias2021, Laguna2021, CARDOSO}. 

Complex network theory is a different approach to studying the dynamics of the economy. It considers economic agents as the nodes, whereas their relations form the edges \cite{Squartini2018,Bardoscia2021, Ma2013, Braunstein2013}. Studies have shown the existence of preferential ties in networks that describe relationships among economic agents \cite{fagiolo-2008, vitali-pone-2011, liu2022preferential}. Homophily --and its counterpart heterophily-- can be understood as deviations of the random assortment of nodes in complex networks \cite{mcpherson2001}. In other words, agents that share characteristics are more likely to bond. \cite{jackson2021} Discusses how homophily is one of the driving forces that increases inequality. According to him, the shortage of connections between communities holds back social capital, information, and opportunities.

Inequalities among ethnicity, gender, and race are examples of divided societies where people from a particular subgroup have more status, power, and wealth than others \cite{berreman,anthias, grusky}. Indeed, in many societies, dominant classes are predominantly composed of people associated with a single group, and economic inequality might be related to this non-economical classification \cite{campante,kilsztajn, shaikh, darity}. This type of stratification may be a result of preferential linkage between persons.

Brazil is one example of how social stratification by race and gender affects economic inequality. According to the Brazilian Institute of Geography and Statistics (IBGE), in 2020 approximately $43\%$ of the country's population are White, and $56\%$ consider themselves Afro-Brazilian (the racial group composed of Black and Brown individuals). Yellow and Indigenous form the other 1\%. %Due to miscegenation, the race data is from self-declaration. 
From $2018$ to $2020$, the average income of Whites remained  $50\%$ larger than the average income of Afro-Brazilians. The disparity is even more considerable between white male and black/brown female incomes, which exceeds $150\%$ \cite{ibge}.

In networks of firms,  homophily could lead to the formation of clusters based on geographic location, where firms within the same geographic region are more likely to be connected than firms in different locations. The geographical homophily in the global market may be the motivation behind the formation of regional trading blocks. It preserves the industry of a region of the globe from international competition \cite{Jilberto1998, frankel}. Analyses of International Trade Networks show other levels of homophily besides location. Richer countries (in terms of Gross Domestic Product) have more trade connections and are more connected among them than poorer ones \cite{garlaschelli-PRL-2004, garlaschelli-PA-2005}. More recently, \cite{liu2022preferential} demonstrated that edges occur more frequently among countries that invest in research and development.  Nonetheless, international trade occurs through firms and companies and not by the countries themselves.  In this sense, to accurately model this phenomenon, it is necessary to take into account the institutional context of countries and trading blocks.

In this work, we extended a recently proposed dynamic network model \cite{kohlrausch2024}, with the addition of a two group interaction. We consider a complex network framework where each agent is a node and their transactions are the edges. The network is dynamic, so the agents are able to create and destroy links depending on their wealth and the group the involved nodes belong, both creation and sustain of links are more likely for wealthier agents. In one group agents prefer the creation of intragroup links, while members of the other are less likely to connect with each other.

\section{Model}\label{secmodel}

The system is divided into two groups, namely $A$ and $B$, each with $N/2$ agents; to each agent we attribute a risk factor ($\alpha$) and a income ($\omega$), both independent random variables within the interval $[0,1)$. 
To create the network, at each group we select $z$ agents and determine that they are fully connected. It is important to note that the parameter $z$ holds no sway over the dynamic or stationary aspects of the model; its sole purpose is to initialize the network. After that, we select a random unconnected agent $j$ to connect with a network agent $i$ with a probability that depends on the group and wealth of both agents
\begin{equation}\label{prob_con}
\left\{
\begin{array}{lc}
 P_{i,j}^{A,B} = [\omega_i(t) +\omega_j(t)]/W & \quad \textrm{if } i\in A, j \in B 
\\% Se i \in B e j \in A, essa equação acima não contempla. Preciosismo meu, sei.
 P_{i,j}^{A,A} = [\omega_i(t) +\omega_j(t)]/W + \delta & \quad \textrm{if } i,j \in A
\\
P_{i,j}^{B,B} = [\omega_i(t) +\omega_j(t)]/W - \delta & \quad \textrm{if } i,j \in B, 
\end{array}
\right.
\end{equation}
where $W=\sum_l \omega_l(t)$ is a sum over all the connected agents and $\omega_i(t)$ is the wealth of agent $i$ at time $t$. The parameter $\delta$ is a term that sets the preference of an agent to connect to another member of the same group. Since $\delta$ add the probability of $A$,$A$ connections while reducing $B$,$B$, our model considers that agents of both groups are more likely to trade with group $A$ at the expense of $B$. For practical purposes we set that 
%($P_{i,j}^{B,B} < P_{i,j}^{A,B} < P_{i,j}^{A,A}~ \forall~ \delta > 0$)
\begin{equation}
    \left\{
    \begin{array}{cc}
    \textrm{if } P_{i,j}\leq 0     & \textrm{then } P_{i,j} = 0  \\
    \textrm{if } P_{i,j}\geq 1     & \textrm{then } P_{i,j} = 1. 
    \end{array}
\right.
\end{equation}

This process continues until the $N$ agents are added to the network. After setting the network, we begin two alternating processes, the exchange of wealth between the connected agents and the rewiring of connections. 

The wealth exchange process does not depend on the agents groups and each agent trades wealth with all its connections. To determine the amount of wealth traded in each transaction ($d\omega$) we considered the yard-sale rule (\cite{Benhur, Hayes})
\begin{equation}
    d\omega = \textrm{min}[\alpha_i\omega_i(t);\alpha_j\omega_j(t)]. 
\end{equation}

To determine the direction wealth flows, we define a probability of the poorer agent winning the transaction, which is given by (\cite{Scafetta})
\begin{equation}\label{prob_exc}
     p_{i,j} = \frac{1}{2} + f \times \frac{|\omega_i(t)-\omega_j(t)|}{\omega_i(t)+\omega_j(t)},
\end{equation}
where $f$ is the social protection factor, which varies between $0$ and $1/2$. The case $f = 0$ corresponds to unbiased trades where each agent has the same probability of winning. The wealth of the agents $i$ and $j$ after the trade will be 
\begin{align*}
 \omega_i(t+1) = \omega_i(t) + d\omega &&  \omega_j(t+1) = \omega_j(t) - d\omega,
 \end{align*}
where the agent $i$ is the winner of the transaction, in this way the total wealth is always conserved. After all the connected agents have traded wealth with all their connections,  the rewiring process takes place.

The rewiring process starts by randomly selecting a pair $i',j'$ of agents, if this pair is disconnected, the probability of creating a new connection follows Eq.\ref{prob_con}. If the selected pair is already connected, there is a probability $Q_{i',j'} = 1 - P_{i',j'}$ of breaking the connection. In order for all agents to be able to participate in this process, we select $N/2$ random pairs of agents to rewire their connections. After this process is completed, we go back to the wealth exchange process respecting the new network that has been formed. In this way, the evolution of the system depends on the wealth exchange and rewiring processes, which are not independent. 

We use a MCS (acronym for {\it Monte Carlo Step}) as the unit of time, which is defined as the time necessary to make all the wealth exchanges and rewire the $N/2$ connections. We perform simulations for systems with size $N=10^3$ and the results are averaged over $10^3$ independent samples which evolve until the equilibrium is reached.

\section{Results}

\begin{figure}
    \centering
    \includegraphics[width=1\textwidth]{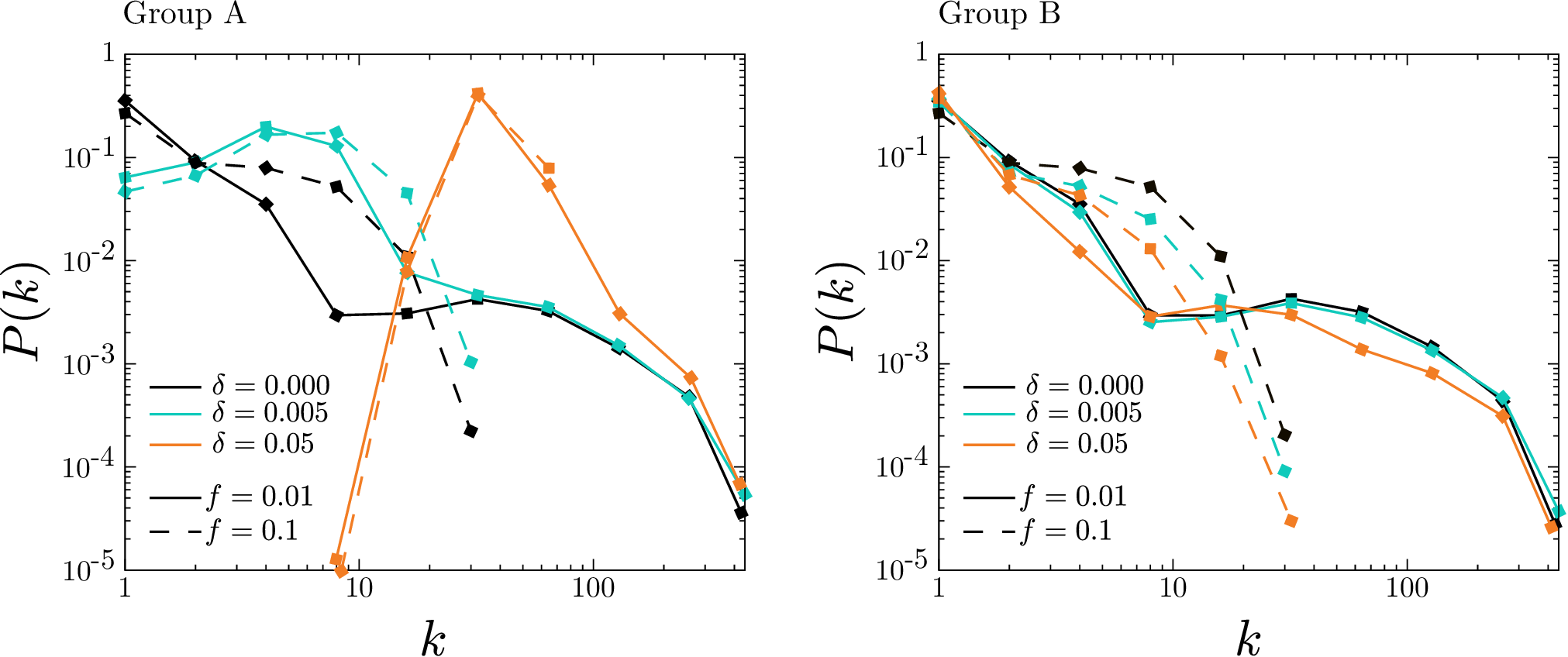}
    \caption{Degree distribution for groups $A$ (left) and $B$ (right) with different values of $\delta$ with $f=0.01$ and $0.1$, solid and dashed lines respectively.}
    \label{degdist}
\end{figure}

We start by showing the degree distribution for each of the groups (Fig~\ref{degdist}). In group A (left panel), we see that in the $f=0.01$ case there is a set of highly connected agents ($k>100$) for all analyzed values of $\delta$. This set of agents disappears for $f\geq 0.1$ and does not appear even when A is highly homophilic ($\delta=0.05$). Small values of homophily ($\delta =0.005$) rises the connection probability of low wealth agents, so a growth in $P(k)$ is perceived for $k=[2,16]$ while the end of the distribution remains unchanged. Large values of $\delta$ change the whole shape of this distribution by creating a extremely connected group, where there are no agents with $k<8$. In this sense, in group A, $\delta$ has the effect of producing a vast number of links, while $f$ redistributes these. 
In group B (Fig~\ref{degdist} right panel) we also observe a framework of highly connected agents for $f=0.01$, which is broken by the increase in social protection. The increase in $\delta$ presents a decrease in $P(k)$ for all values of $k$, which is more pronounced for $f=0.1$.

\begin{figure}
    \centering
    \includegraphics[width=1\textwidth]{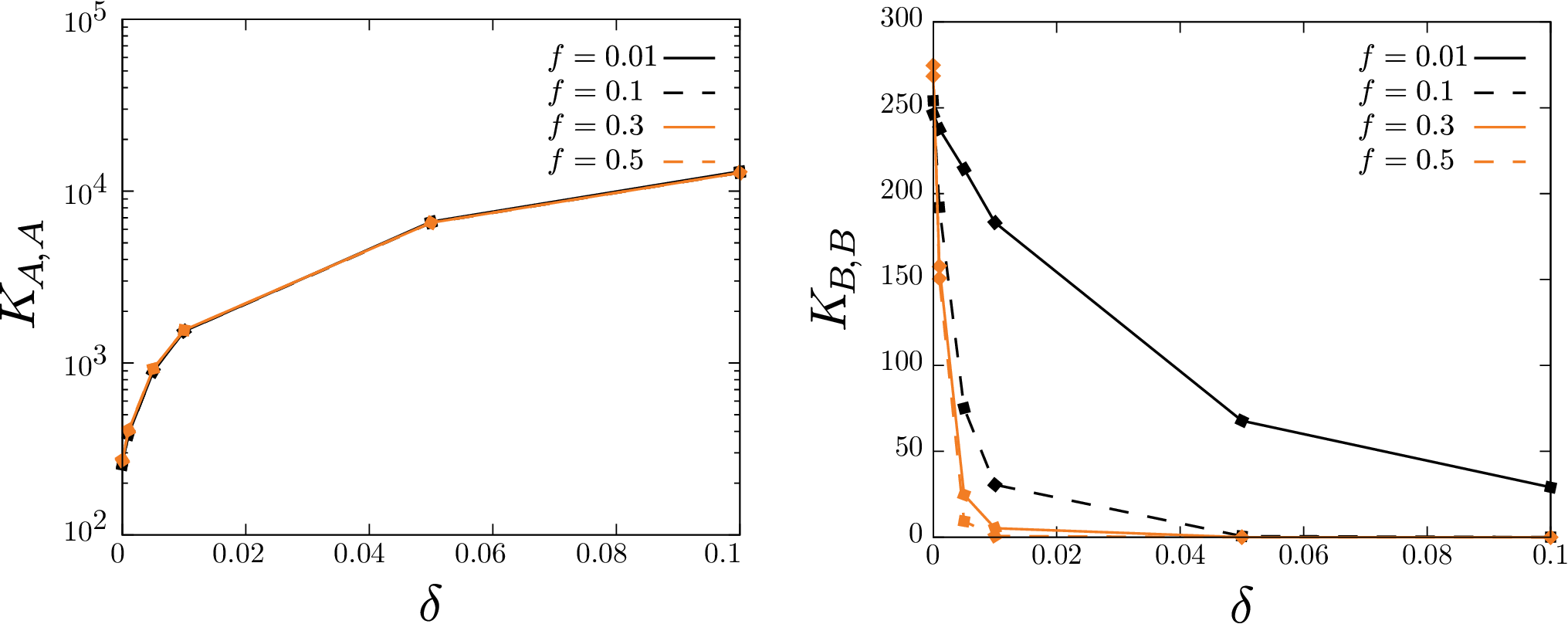}
    \caption{Total intragoup connections in groups $A$ (left) and $B$ (right) as a function of $\delta$ for different values of social protection.}
    \label{K_AA_BB}
\end{figure}

The total intragroup connections ($K_{A,A}$ and $K_{B,B}$) of each group is shown in Fig~\ref{K_AA_BB}. In group A (left panel),  $K_{A,A}$ presents a vast increase with $\delta$, going from near $250$ connections to over $10 000$. $K_{A,A}$ also increases with $f$, however, this growth is very little pronounced, especially compared to the effects of $\delta$. The links between agents of group B ($K_{B,B}$) rapidly decreases with the increase of $\delta$ (Fig~\ref{K_AA_BB} right panel), and the only case when this groups remains connected is for $f=0.01$.  Nevertheless, even with the very low quantities of intragroup interactions when $\delta$ increases, the degree distribution present the same patterns for low and high values of this parameter (Fig~\ref{degdist}). In this way, the wealth dynamics of the intergroup interactions seems to be the main driving force of the agents connection, and a strong $\delta$ can make the $B$ group itself disconnected, but cannot isolate its agents, as they remain interacting with the group $A$.

\begin{figure}
    \centering
    \includegraphics[width=1\textwidth]{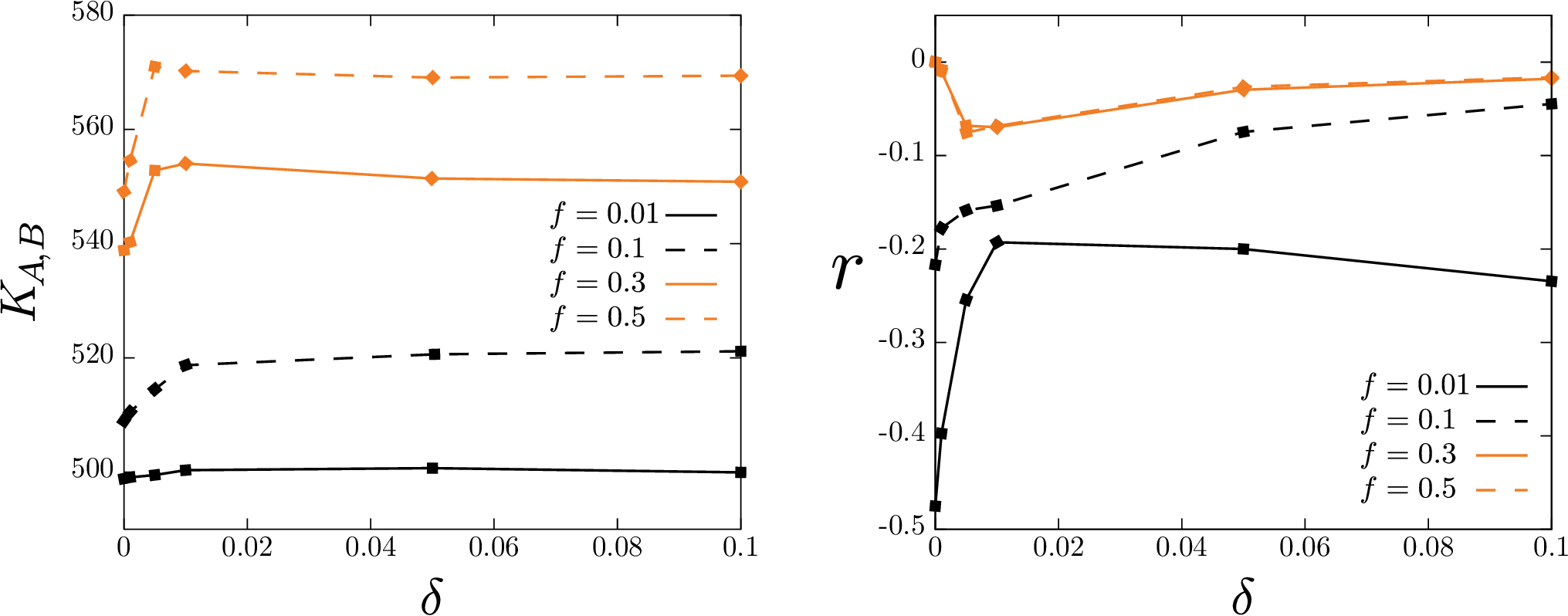}
    \caption{Total intergroup connections (left) and assortativity (right)  as a function of $\delta$ for different values of social protection.}
    \label{assor}
\end{figure}

In Figure~\ref{assor} left panel, is presented the total interchanged connections between groups ($K_{A,B}$). Although $\delta$ makes $A$,$B$ connections preferable to $B$,$B$ ones, the social protection factor ($f$) has a much greater influence in $K_{A,B}$. Since $\delta$ is not directly included in the probability of connection between distinct groups (Eq~\ref{prob_con}), the increase in $K_{A,B}$ that happens for small values of $\delta$ must come from a change in the wealth distributions. Furthermore, the increase in $K_{A,B}$ brought by $f$ can be related to a more egalitarian distribution of wealth, where more agents are able to participate in the $A$,$B$ exchanges.

Aiming to gather information on how the agents are connected with each other, our next step will be investigating the network assortativity ($r$), which measures the Pearson correlation of the degree of connected nodes \cite{Newman2002}. This coefficient can be expressed in a operational way as

\begin{equation}
  r = \frac{M^{-1}\sum_i^M j_i k_i - [M^{-1}\sum_i^M \frac{1}{2}(j_i +k_i)]^2}{M^{-1}\sum_i^M \frac{1}{2}(j_i^2 +k_i^2) -[M^{-1}\sum_i^M \frac{1}{2}(j_i +k_i)]^2 },
    \end{equation}

where the sum in $i$ is made on the edges of the network, $M$ is the total number of connections, $j_i$ and $k_i$ are the degrees of the agents at the end of the $i$ edge. This coefficient varies in the interval $-1\leq r \leq 1$. For $r<0$ the network is  disassortative and the highly connected nodes are linked to the ones with low degrees. For $r>0$ the network is assortative and the nodes are connected with others with similar degree. When $r=0$ the network is called non-assortative and there is no correlation between connected agents. For the assortativity of individual groups ($r_A$ and $r_B$) we disconsider intergroup links in the sum of $M$, however, we still consider the total degree of the agents, \textit{i.e.}, including intergroup connections.

\begin{figure}
    \centering
    \includegraphics[width=1\textwidth]{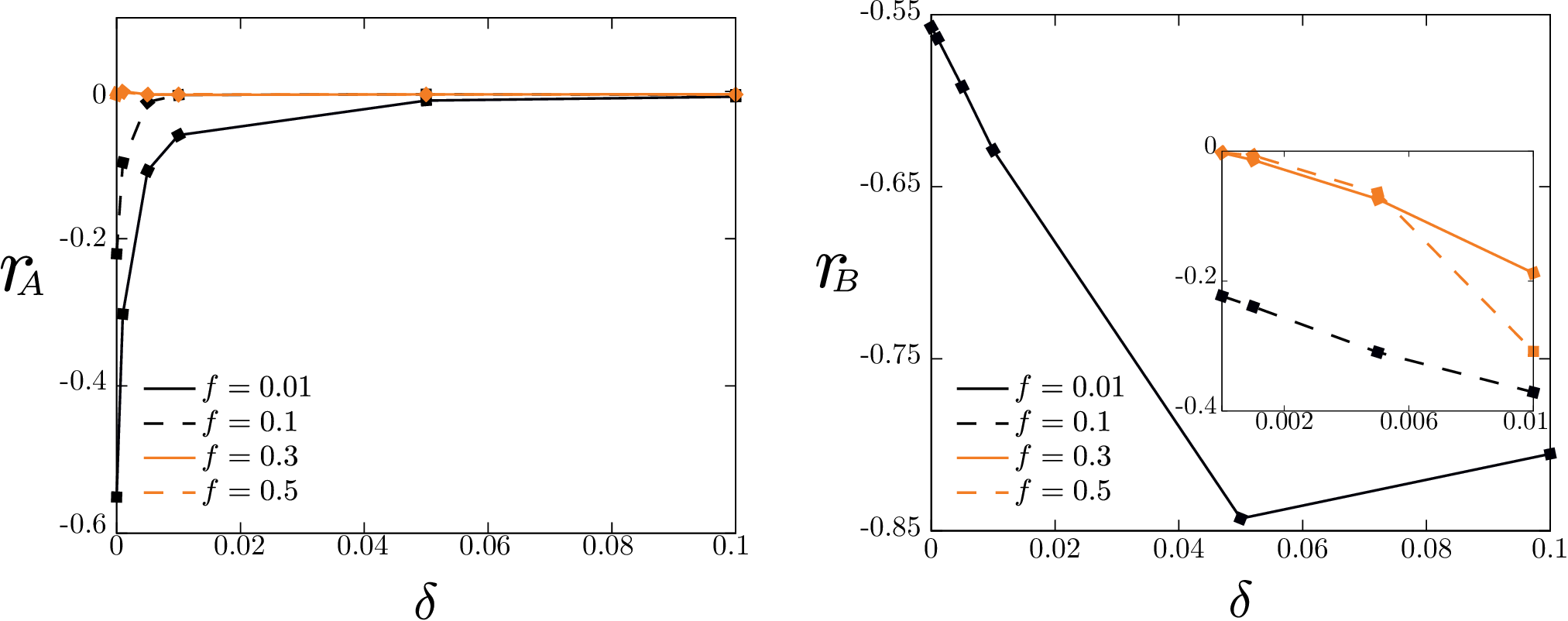}
    \caption{Assortativity of groups $A$ (left) and $B$ (right) as a function of $\delta$ with different values of social protection.}
    \label{assor_AB}
\end{figure}

The group $A$ is disassortative for small values of $f$ and $\delta$ (Fig~\ref{assor_AB} left panel), besides their different effects on the degree distribution, the increase on both these parameters lead to a decrease in the assortativity magnitude. In this sense, when $f\geq 0.3$ or $\delta\geq0.05$ the $A$ network becomes non-assortative. The disassortative behavior in lower values of $f$ and $\delta$ demonstrate that the high degree are connected with a vast number of low degree agents, acting as hubs in the network. This way, in group $A$ an increase in $\delta$ rises the probability of connection for the low degree agents, thus leading to the growth of $K_{A,A}$ and consequently the shrinkage of $|r_A|$. Furthermore, the growth of $f$ eliminates the hubs in the network by creating a more equal degree distribution.

The group $B$, while it remains connected, is disassortative for all the values of $f$ and $\delta$ studied (Fig~\ref{assor_AB} right panel). The increase of $\delta$ makes the connection between the poorest $B$ agents unlikely, thus, the values of $|r_B|$ enhance as these mostly connect with the richest. In the inset of Fig~\ref{assor_AB} we present the assortativity of group $B$ for $f\geq0.1$, as in this case the group is disconnected for $\delta>0.01$ we only show results until this value. As we can see, $|r_B|$ shrinks with increasing $f$, becoming very close to zero for $\delta\leq0.005$ and $f\geq0.3$. In this sense, the effect of the social protection in this group is the same as in $A$, generates a more equal degree distribution which diminishes its disassortative characteristics. Also, we perceive a slight decrease in $|r_B|$ going from $\delta=0.05$ to $\delta=0.1$ when $f=0.01$. Since in this regime of $f$ there is the presence of hubs in the network, this large value of $\delta$ seems to break some connections of this highly connected agents with the lower degree ones from their group, diminishing $K_{B,B}$ and $|r_B|$.

Looking at the assortativity of the whole network (Fig~\ref{assor} right panel), not surprisingly, we found that  $f$ diminishes $|r|$. However, we only found $r=0$ when $\delta=0$, for values of $\delta>0$, even for strong social protection, $r$ is always less than zero. Furthermore, the $\delta$ effects in this quantity are quite more complicated, and depend on the value of $f$. Starting with the $f=0.01$ scenario, we observe that $|r|$ rapidly decreases for small increases in $\delta$, then there is a turning point for $\delta = 0.01$ and $|r|$ presents a slight increase. Considering the great values of $K_{A,A}$ when compared with $K_{A,B}$ and $K_{B,B}$, the connections between $A$ agents must be the main driving force in reducing $|r|$. However, for $\delta>0.01$, $r_A$ is very small, therefore the increase in $|r|$ after this point must come from a disassortative characteristic in the connections between groups. In this way, the highly connected agents of each group act as hubs also in the other group. For $f\geq0.1$, $|r|$ always decreases after $\delta=0.01$ following the behavior of $r_A$. However, the $AB$ connections still maintain a small disassortative behavior to the network. In the same sense, these intergroup connections are able to bring $r$ to negative values when $\delta$ is small for $f=0.3$ and $0.5$, where $K_{A,B}$ is nearly half of $K_{A,A}$. However, going to larger values of $\delta$, $K_{A,A}$ substantially increases, pushing $r$ towards zero.

\begin{figure}
    \centering
    \includegraphics[width=1\textwidth]{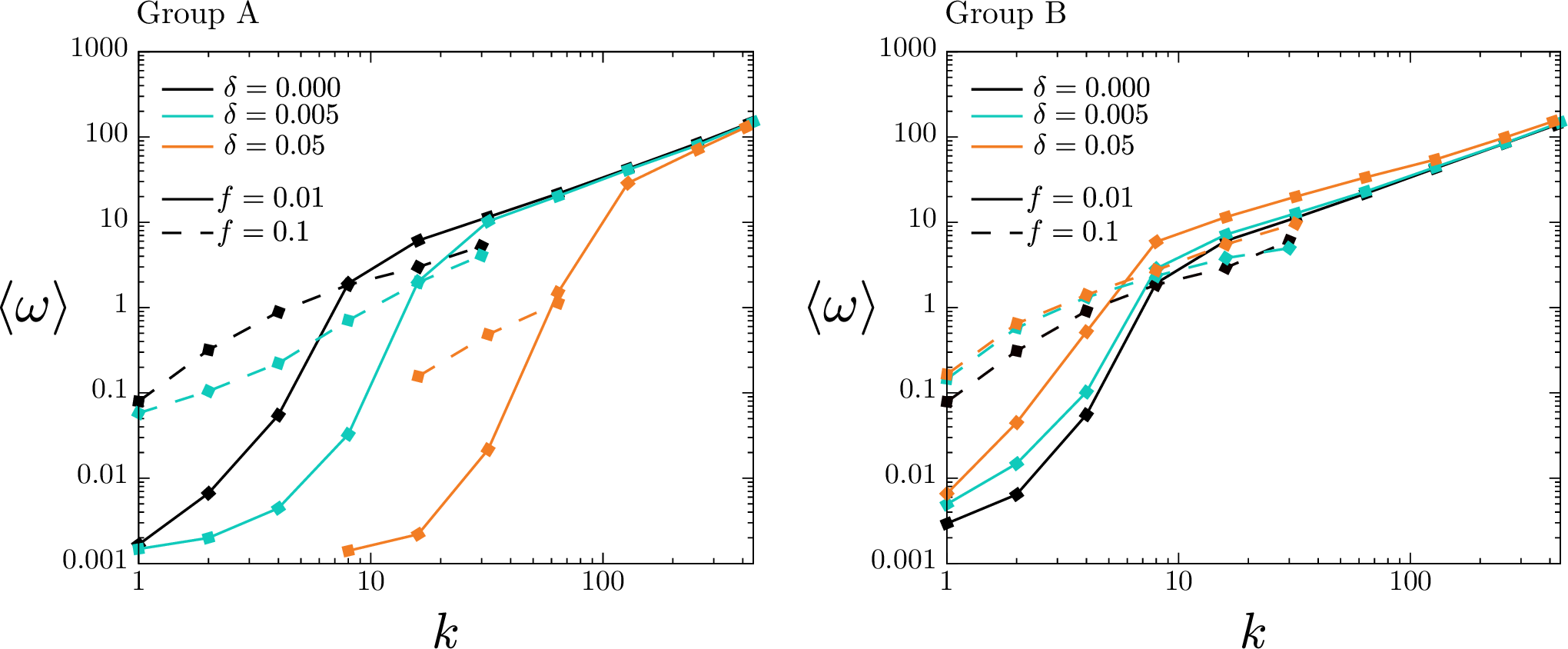}
    \caption{Mean wealth of agents with degree $k$ for groups $A$ (left) and $B$ (right) with different values of $\delta$ with $f=0.01$ and $0.1$, solid and dashed lines respectively.}
    \label{wealth_degree}
\end{figure}

We now start by looking at how wealth is distributed in the network. For this matter, in Figure~\ref{wealth_degree} we present the mean wealth of agents with degree $k$ for groups $A$ (left panel) and $B$ (right panel), respectively. In both groups it is observed that for $f=0.01$ and $0.1$ the wealth of an agent increases with their degree. Nevertheless, wealth and degree inequalities are significantly attenuated when $f$ increases from $0.01$ to $0.1$. Actually, for $f=0.01$ the richest agents have $\langle\omega\rangle >100$ and $k>400$, while for $f=0.1$ this values are reduced to $\langle\omega\rangle<10$ and $k\leq 64$. 
The effects of $\delta$ are opposite in the two groups. In $A$, an increase of $\delta$ diminishes $\langle \omega \rangle$ for a given $k$ (Fig~\ref{wealth_degree}). In this context, $\delta$ creates more intragroup connections with the poor agents, decreasing the mean values of wealth and dislocating the $\langle \omega \rangle \times k$ distribution towards larger degrees. On the other hand, in $B$ (Fig.~\ref{wealth_degree} right), $\delta$ destroys the intragroup connections with the poor, thus $\langle \omega\rangle$ increases with $\delta$. The changes caused by $\delta$ are quite more notable in $A$ than in $B$, which can be explained by the significant increase in $K_{A,A}$ with $\delta$. Also, the degree distribution of $B$ (Fig~\ref{degdist} right) does not present such pronounced changes as for $A$ (Fig~\ref{degdist} left).

\begin{figure}
    \centering
    \includegraphics[width=1\textwidth]{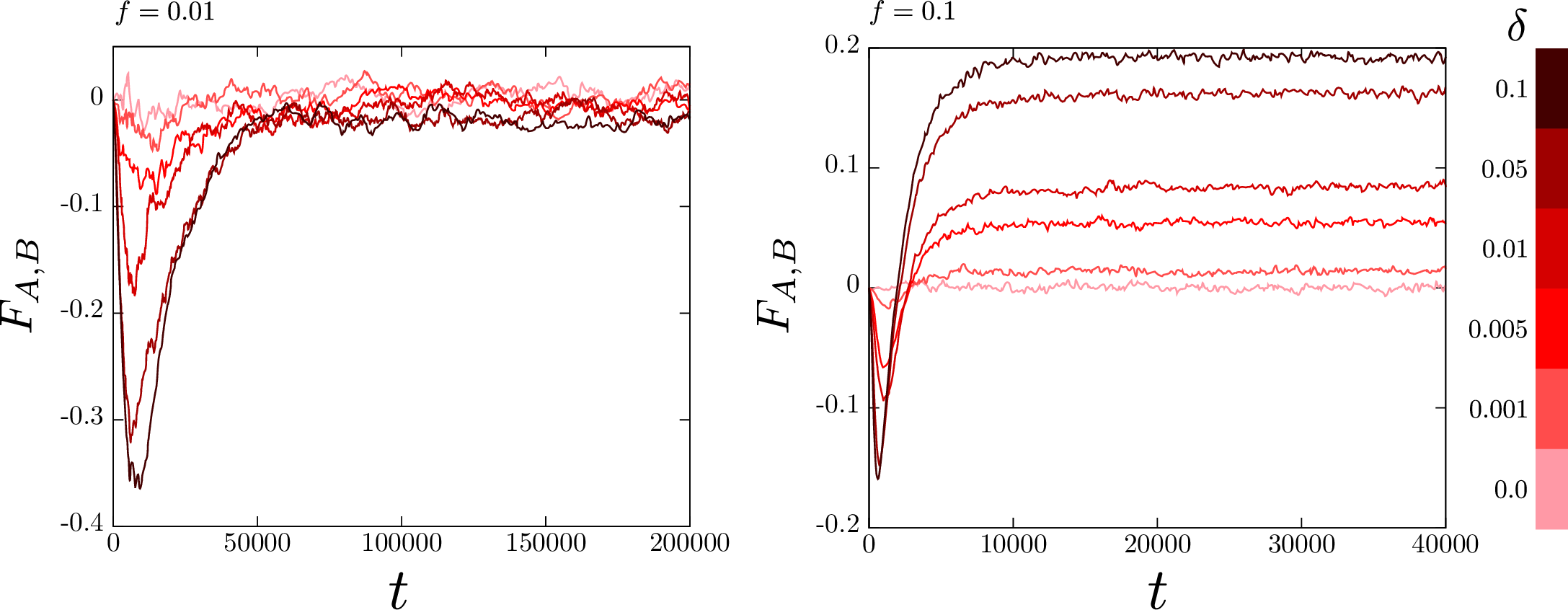}
    \caption{Net wealth flow between groups for different values of $\delta$ with $f=0.01$ (left) and $0.1$ (right) as function of time.}
    \label{0.01_0.1_dif}
\end{figure}

In Figure~\ref{0.01_0.1_dif}  we show the temporal behavior of the net wealth flow between groups ($F_{A,B}$), 
\begin{equation}
    F_{A,B} = (\Omega_A(t) - \Omega_B(t))/\Omega,
\end{equation}
where $\Omega_{A(B)} = \sum_{i\in A(B)}\omega_i(t)$  is the total wealth in group $A$($B$) and $\Omega$  is the wealth in the system, which is constant, for $f=0.01$ (left panel) and $0.1$ (right panel), with different values of $\delta$. As the two groups continue to trade wealth, the net exchange always varies in time. However, for sufficiently large times, this variations occur around a fixed point, which we define as the equilibrium state. For $f=0.01$, the system takes a much longer time to reach the steady state, presenting an extended transient behavior where a significant fraction of wealth flows to group $B$, then returns to $A$. In this transient, the wealth initially gained by the group B increases with $\delta$, reaching $35\%$ of the total wealth at its maximum when $\delta=0.1$. Although we observe the same transient behavior for $f=0.1$, it happens in a much shorter time and reduced magnitude.

\begin{figure}
    \centering
    \includegraphics[width=1\textwidth]{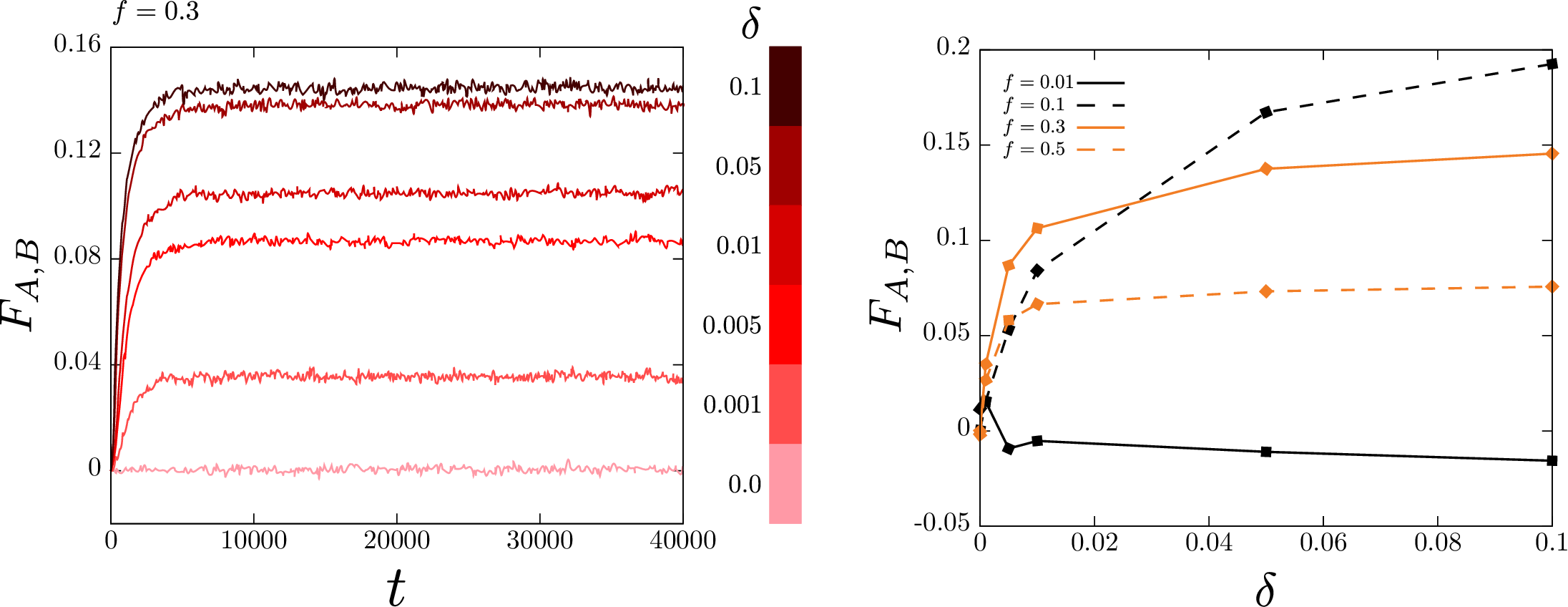}
    \caption{Left: net wealth flow between groups for different values of $\delta$ with $f=0.3$ as function of time. Right: net wealth flow between groups and in the steady state as a function $\delta$ for different values of social protection.}
    \label{dif_vs_d}
\end{figure}

In Figure~\ref{dif_vs_d} we present $F_{A,B}$ for $f=0.3$ (left panel) and in the steady state for different $f$ (right panel). For $f=0.3$ we observe  the absence of the $F_{A,B}$ transient beahvior, where this quantity rapidly stabilizes even for high values of $\delta$. Looking to $F_{A,B}$ steady state (Fig~\ref{dif_vs_d} right),  in the $f=0.01$ case, group $B$ extracts a small fraction of wealth from $A$, which increases with $\delta$. This result can be related with the transient behavior presented in Figure~\ref{0.01_0.1_dif}, suggesting that a slight percentage of wealth gained by $B$ in the early stages of the simulation is kept by its agents. Nevertheless, in very small values of $\delta$ there is a wealth flow to $A$. For values of $f\geq 0.1$ we observe a increasing wealth flow to group $A$ with $\delta$. Increases of social protection diminishes the economic disparity between groups, which is expected as $f$ favors a more egalitarian distribution. However, the opposite scenario occurs for small values of $\delta$, where $F_{A,B}$ increases as we go from $f=0.1$ to $0.3$. This unusual result for $f=0.1$ can also be related to the transient behavior, which is not present in stronger social protection.

\begin{figure}
    \centering
    \includegraphics[width=1\textwidth]{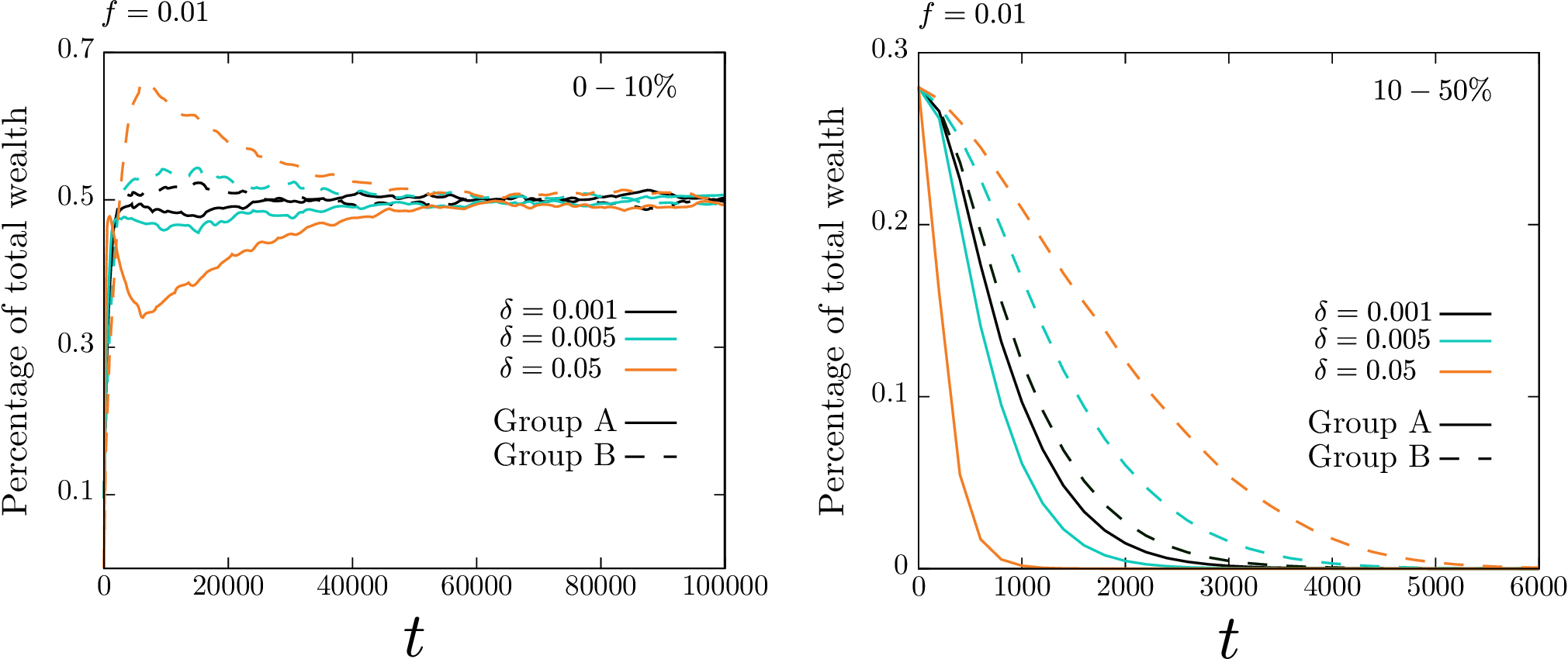}
    \caption{Percentage of the total wealth held by the high (left) and middle (right) classes as a function of time with different $\delta$ for $f=0.01$ in groups $A$ and $B$, solid and dashed lines respectively.}
    \label{percentis_0.01}
\end{figure}

\begin{figure}
    \centering
    \includegraphics[width=1\textwidth]{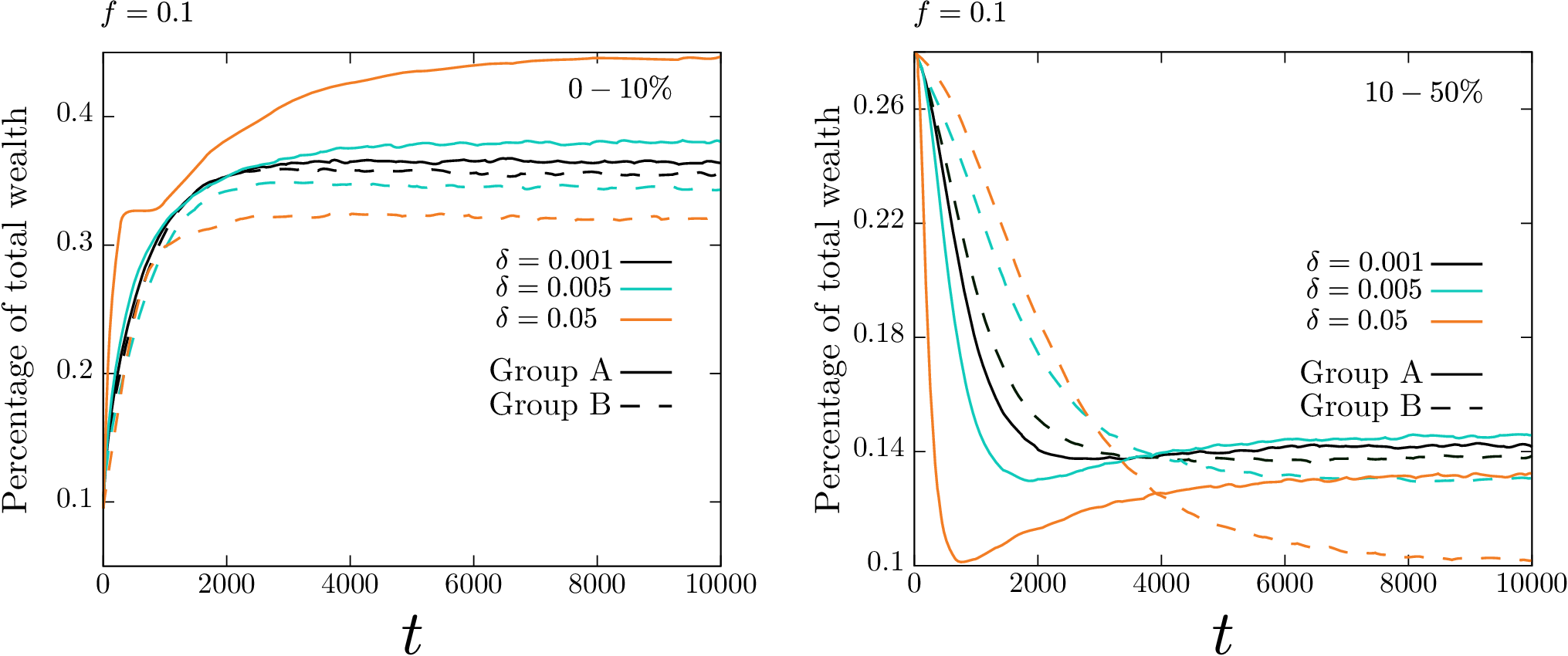}
    \caption{Percentage of the total wealth held by the high (left) and middle (right) classes as a function of time with different $\delta$ for $f=0.1$ in groups $A$ and $B$, solid and dashed lines respectively.}
    \label{percentis_0.1}
\end{figure}

\begin{figure}
    \centering
    \includegraphics[width=1\textwidth]{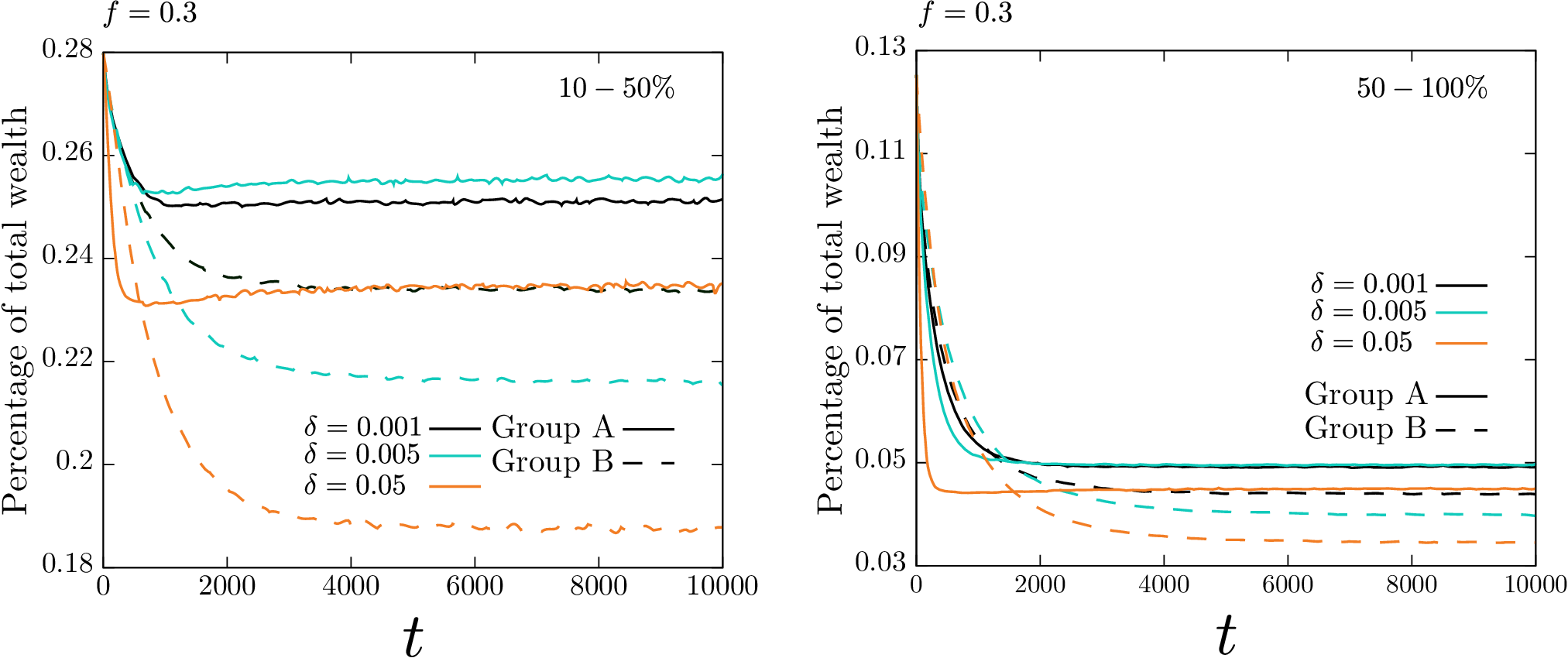}
    \caption{Percentage of the total wealth held by the middle (left) and bottom (right) classes as a function of time with different $\delta$ for $f=0.3$ in groups $A$ and $B$, solid and dashed lines respectively.}
    \label{percentis_0.3}
\end{figure}

To better understand the wealth flow between groups and how its distributed in these, we divide each group in four different classes. The top $10\%$  will be referred as the high class. The middle class will be designed as the $10-50\%$ richest agents. For the last, the poor agents are the bottom $50\%$ of the wealth distribution. We present in Figures~\ref{percentis_0.01}, \ref{percentis_0.1} and \ref{percentis_0.3} how these groups evolve over time for different values of $f$ and $\delta$. We will make this analysis looking at the percentage of the total wealth of the system held by each class. 

We show in Figure~\ref{percentis_0.01} the evolution of the high and middle classes, left and right panel respectively, for $f=0.01$. We see that the wealth flow to group $B$ presented in Figure~\ref{0.01_0.1_dif} is directed to its wealthier agents. In the transient, the wealthier agents of $A$ lose a significant amount of wealth, which is recovered afterwards. However, this loss alone cannot explain all the wealth gained by group $B$. In this sense, the rich $B$ agents must also extract wealth from other $A$ classes. What we observe, in both groups, is a bankruptcy of the bottom and middle classes, besides the last one takes a longer time to reach $\omega = 0$. The $\delta$ parameter affects this dynamic by accelerating the bankruptcy of $A$ agents and decelerating on group $B$. As $\delta$ diminishes the intragroup $B$ connections, especially with lower wealth agents, it can protect the $B$ middle and bottom classes from intragroup losses of wealth. However, in a long run, top classes are able to extract all the wealth from the bottom ones. In this way, in weak social protection, the wealthier agents of each group are able to gain wealth from the poor ones of the other group. In a short time, the agents $B$ seem to have an economic advantage over $A$, which is not perceived in the equilibrium state.

In Figure~\ref{percentis_0.1} we present the evolution of different classes for $f=0.1$. This increase in social protection rapidly decreases the percentage of wealth held by the top $10\%$ and prevents the bankruptcy of middle class. The half poorest still go to the $\omega=0$ state in the equilibrium and present the same effect of $\delta$ in the dynamic. Nonetheless, $\delta$ has a much greater influence in the equilibrium state for other percentiles. In group $A$ we perceive a great gain of wealth by the high class as $\delta$ goes from $0.005$ to $0.05$, showing that the jump in $F_{A,B}$ presented in Figure~\ref{dif_vs_d} is referring to the top percentiles. On the other side, the group $B$ high class present a loss of wealth as $\delta$ increases. As in this case there almost none intragroup connections in $B$ and a very small increase in intergroup links (see Fig~\ref{assor} left and \ref{K_AA_BB} right panels), the more homophilic characteristics of $A$ agents favour them against $B$. Looking at the evolution of the middle class in these groups we still see that, in short times, $\delta$ accelerates the losses of wealth in $A$ while protecting the $B$ agents. When the system goes to equilibrium, this dynamic is reversed as $A$ agents recover part of their wealth, giving rise to the $F_{A,B}$ transient, and the $A$ middle class ends up richer than its $B$ analogue.  In this sense, in the $f=0.1$ case, the increase in $K_{A,A}$ creates a more participatory middle class in the economic exchanges, as the social protection prevents its bankruptcy. Thus, the $10-50\%$ $A$ class is able to sustain a richer high class than in $B$, enabling a bigger wealth flow to the homophilic group.

In the case of $f=0.3$ (Figure~\ref{percentis_0.3}), the bottom class holds a more significant amount of wealth, which diminishes with a  $\delta$ increase.  This comes from the redistribution of wealth given by the increase in social protection, passing the money held by the high class when $f=0.1$ to the bottom ones, especially to the $10-50\%$ percentile. On another way, even that $\delta$ seems to favour the $A$ group by raising the total amount of money on it, the more homophilic this group is, more unequal he becomes, as we see the middle and bottom classes losing wealth with $\delta$. Thus, the increase of $\delta$ means a disadvantage to agents in the $10-100\%$ percentiles of both groups. Furthermore, the same classes of $B$ perceive much larger losses of wealth with $\delta$, what can be clearly seen in the $10-50\%$ percentile. Nevertheless, an increase in the $A$ middle class wealth is perceived from $\delta=0.001$ to $0.005$, accompanied by a more significant loss in $B$. Larger values of $\delta$ shrink the $A$ $10-50\%$, suggesting that weak homophilic characteristics can favour middle class agents.

\section{Conclusions}\label{secconclu}

In this work, we introduced an complex network agent-based model with two interacting groups, namely $A$ and $B$. The two groups differ in a parameter $\delta$ that increases the probability of connection between agents belonging to group $A$, and decreases in group $B$ (Eq.\ref{prob_con}). In this sense, this parameter tries to a mimic a homophilic behaviour in the first group. In turn of that, as $\delta$ increases, $B$ agents are more likely to connect with those in $A$ than with their own group. So, this parameter acts as a greater economic appeal to group $A$. This way, we relate the $\delta$ increase with the formation of stronger alliances between countries participating of a given trading block, such as the MERCOSUR or the European Union. In this analogy, the agents of each group represent firms or companies inside ($A$) and outside ($B$) the trading block. Thus, as the economic trades inside a given trading block intensify, represented by an $\delta$ increase, it gets more appealing to foreign investors, denoted by the $B$ agents preference for $A$. Another way to look at our model, is to understand it as a single society constituted of two groups, in which one of those, the group $A$, segregates the other. This phenomena is related to homophily in social networks, and is pointed out to be one of the mechanisms that aggravates economic inequalities \cite{jackson2021,mcpherson2001}.

Aside from the homophilic parameter, the phase space is also constituted  by a social protection factor $f$, which favors the poorest agent in every wealth exchange (Eq.\ref{prob_exc}). The $f$ parameter was already studied in mean field \cite{Benhur,Iglesias2021,Nener2021} and dynamic complex network \cite{kohlrausch2024} works, where it is related to governmental social programs or infrastructure investments. In the context of international trades, $f$ is a way to mimic different regulations on trades, what remains true for a closed country. To define the amount of wealth traded, we consider the yard-sale rule \cite{Hayes}, in which the two agents participating in the transactions bet a fraction of their wealth defined by a risk factor $\alpha$. The model then alternates in two processes, the rewire of connections in the complex network and the wealth exchange between all the linked agents. As the probability of connection depends on the agents wealth, these two process are dependent of each other. We investigated the model on values of $f$ ranging from $0$, where there is no favoring of the poor, to $0.5$, and $\delta$ from $0$ to $0.1$, where the group $B$ is completely disconnected for almost all values of $f$.

 In the low protection scenario ($f=0.01$), the wealth concentrates in just a few agents, which appears as hubs in the network. For $f=0.01$ and $0.1$ we observe a transient behavior, where the $B$ group is able to extract some wealth from $A$. This transient is intensified by $\delta$ and its related to the difficulty of the poor $B$ agents to create connections in the early stages of the simulation. Thus, as they remain disconnected they are not able to neither gain or lose wealth, and $\delta$ acts protecting them from intragroup losses. Regardless of that, its important to mention that in a real situation it is a clear disadvantage to be completely disconnected, since economics trades are fundamental to generate production, what is not yet accounted in our model. Nevertheless, when $f=0.01$ the $10\% A$ richest agents are able to recover wealth from $B$, and the two groups end with the same wealth. In this sense, the small social protection atenuates the $\delta$ parameter influence on the steady state and the economic advantage of being rich presents a stronger effect than the homophily. This economical advantage is accompanied by a topological dominance, thus the richest keep getting richer and the system goes to a extremely unequal scenario, characterized by all the wealth concentrated in the top $10\%$ (Fig~\ref{percentis_0.01}). Although,  we recover the middle class when $f=0.1$ (Fig.~\ref{percentis_0.1}) and the bottom when $f=0.3$ (Fig.~\ref{percentis_0.3}).

When the social protection is strong enough ($f\geq0.1$) the effects of the homophilic parameter in the equilibrium can be clearly seen. As $\delta$ grows, there is an increasing wealth flow from group $B$ to $A$. Even though $\delta$ raises the wealth in $A$ it also makes this group more unequal. Hence, even when $f=0.3$ the wealth gained by this group is not equally distributed, and the richer agents take more advantage of the homophilic characteristics. Nevertheless, a small increase in the wealth of the middle and bottom $A$ classes is perceived when going from $\delta=0.001$ to $0.005$, indicating that a more connected group can favor those agents if there is a strong social protection preventing them from wealth losses. On the other hand, the social protection makes the whole system more egalitarian, and the wealth flow to $A$ diminishes as $f$ increases.

To sum up, we explored a recently proposed dynamic network model \cite{kohlrausch2024} with an additional homophilic parameter. Our results show  a very rich phenomenology that is sensible to small changes in both parameters. In our point of view, besides the simplicity of the model, it is able to reproduce some characteristics of international trade networks, such as the disassortative characteristics of the network \cite{liu2022preferential}. In the context of a closed society, this model is able to reproduce a segregation of the poorer group, characterized by a wealth flow to a more homophilic one.  Nevertheless, high values of $f$ and $\delta$ seems to be distant from real observations. It is important remember that the connections between agents considered here are only of economical nature, disregarding social relations. Furthermore, we believe that more studies in this model considering some characteristics that were excluded here, like taxation and production of wealth, are of great interest.

\bibliography{main.bbl} % Please set the right name for your bib file

%%%%%%%%%%%%%%%%%%%%%%%%%%%%%%%%%%%%%%%%%%%%%%

\end{document}